\documentclass[aps,superscriptaddress,preprint,nofootinbib]{revtex4-1}
\usepackage{graphicx}
\usepackage{color} % for colored text
\def\be{\begin{equation}}
\def\ee{\end{equation}}
\def\ba{\begin{eqnarray}}
\def\ea{\end{eqnarray}}

\begin{document}

\title{Holographic Thermal Relaxation in Superfluid Turbulence}

\author{Yiqiang Du}
\email{duyiqiang12@mails.ucas.ac.cn}
\affiliation{School of Physics, University of Chinese Academy of Sciences,
Beijing 100049, China}

\author{Chao Niu}
\email{niuc@ihep.ac.cn}
\affiliation{Institute of High Energy Physics, Chinese Academy of Sciences, Beijing 100049, China}

\author{Yu Tian}
\email{ytian@ucas.ac.cn}
\affiliation{School of Physics, University of Chinese Academy of Sciences,
Beijing 100049, China}
\affiliation{State Key Laboratory of Theoretical Physics, Institute of Theoretical Physics, Chinese Academy of Sciences, Beijing 100190, China}

\author{Hongbao Zhang}
\email{hzhang@vub.ac.be}
\affiliation{Department of Physics, Beijing Normal University, Beijing 100875, China}
\affiliation{Theoretische Natuurkunde, Vrije Universiteit Brussel and
The International Solvay Institutes, Pleinlaan 2, B-1050 Brussels, Belgium}

\begin{abstract}
Holographic duality provides a first-principles approach to investigate real time processes in quantum many-body systems, in particular at finite temperature and far-from-equilibrium. We use this approach to study the dynamical evolution of vortex number in a two-dimensional (2D) turbulent superfluid through numerically solving its gravity dual. We find that the temporal evolution of the vortex number can be well fit statistically by two-body decay due to the vortex pair annihilation featured relaxation process, thus confirm the previous suspicion based on the experimental data for turbulent superfluid in highly oblate Bose-Einstein condensates. Furthermore, the decay rate near the critical temperature is in good agreement with the recently developed effective theory of 2D superfluid turbulence. %Remarkably, such a decay pattern dominates the relaxation process from a very early time on, which goes beyond the existing theoretical analysis and needs further experimental confirmation.
\end{abstract}

\maketitle

\section{Introduction}
Quantized vortices are topological objects in a 2D superfluid, where vortex-antivortex pairs play a crucial role in various superfluid phenomena, in particular the fascinating quantum turbulence in superfluids. Amazingly, vortex dipole dynamics can nowadays be experimented in a controllable way in atomic Bose-Einstein condensate systems\cite{NSBDA,FBKLH}. In particular, %not only has the thermal activation of vortex pairs been observed in quasi-2D Bose gases\cite{HKCBD,CSS}, but also 
most recently the thermal relaxation process of turbulent superflow has been experimentally investigated at finite temperatures of order the critical temperature, where the vortex pair annihilation is observed and suspected to account for the two-body decay behavior of vortex number after the inevitable drift-out effect has been subtracted for the real experimented sample by hand\cite{KMCSS}.

Compared to these significant experimental developments, our theoretical understanding of dynamics of these quantized vortices is still limited because at {\it finite} temperature the effective dissipative hydrodynamical description for normal fluids does not work in the presence of quantized vortices and all the conventional approaches rely on some phenomenological models, which nevertheless have significant shortcomings. With this in mind, any {\it ab initio} theoretical framework would be greatly desirable. Gratefully, holographic duality provides us with such a satisfactory theoretical framework, in which a complete description of a strongly coupled quantum many-body system, valid at all scales, can be encoded in a classical gravitational system with one extra dimension. In particular, the superfluid at finite temperature is dual to a hairy black hole in the bulk and the dissipation mechanism is naturally built in the bulk in terms of excitations absorbed by the hairy black hole. Thus it allows a first-principles investigation of vortex dynamics by using the dual gravity description of superfluid phase. It is recently shown by holography in the seminal work \cite{ACL} that although the 2D turbulent superfluid kinetic energy spectrum obeys Kolmogorov $-5/3$ scaling law as it does for turbulent flows in normal fluids, the superfluid turbulence demonstrates a direct energy cascade towards a short-distance scale set by the vortex core size, in stark contrast to the hydrodynamical argument for the inverse energy cascade of 2D normal fluid turbulence due to the conservation of enstrophy, which is violated in a superfluid by vortex pair annihilation anyhow.

Inspired by the above sharp results derived from the holographic principle as well as the aforementioned experimental investigation of thermal relaxation for superfluid turbulence, in this paper we shall make such a holographic duality contact closer with the experimental data by initiating a quantitative investigation of temporal evolution of vortex number during the thermal relaxation of superfluid turbulence through numerically solving the coupled nonlinear equations of motion of its gravity dual. Remarkably, not only does our holographic result confirm the suspected vortex pair annihilation induced two-body decay rate, but also the decay rate near the critical temperature is in good agreement with the recently developed effective field theory description of 2D superfluid turbulence in \cite{CL}.\footnote{For the seemingly disparity between our holographic result and the result by effective field theory, please check the conclusion section for a detailed discussion.} We also have reliable results at moderate temperature for the decay rate of vortex number, but more detailed comparison between our holographic simulation and experiments may need more experimental data and new technology due to the drift-out effect and the deviation of oblate Bose-Einstein condensates from truly 2D systems.

\section{Holographic setup}
The simple holographic model for 2D superfluid consists of gravity in asymptotically AdS$_4$ spacetime coupled to a $U(1)$ gauge field $A$ and a complex scalar field $\Psi$ with charge $q$ and mass $m$. The corresponding bulk action is given by\cite{HHH1,HHH2}
\begin{equation}
S=\frac{1}{16\pi G}\int_\mathcal{M} d^4x\sqrt{-g}(R+\frac{6}{L^2}+\frac{1}{q^2}\mathcal{L}_{matter}),
\end{equation}
where $G$ is the Newton's constant,  $L$ is the radius of curvature of AdS, and the matter Lagrangian reads
\begin{equation}
\mathcal{L}_{matter}=-\frac{1}{4}F_{ab}F^{ab}-|D\Psi|^2-m^2|\Psi|^2
\end{equation}
with $D=\nabla-iA$. $\frac{L^2}{16\pi G}\gg1$ is required such that classical gravity is reliable, which corresponds to the large $N$ limit of the dual field systems such as ABJM theory. In addition, we shall work in the probe limit,  namely the matter fields decouple from gravity, which can be achieved by taking the large $q$ limit\cite{large}. One thus can put the matter fields on top of Schwarzschild black brane background, which can be written in the infalling Eddington coordinates as
\begin{equation}
ds^2=\frac{L^2}{z^2}(-f(z)dt^2-2dtdz+d\mathbf{x}^2),
\end{equation}
where the blackening factor $f(z)=1-(\frac{z}{z_h})^3$ with $z=z_h$ the location of horizon and $z=0$ the AdS boundary. The behavior of matter fields is controlled by the equations of motion in the bulk as
\begin{eqnarray}
D_aD^a\Psi-m^2\Psi=0,\quad\nabla_aF^{ab}=i(\bar{\Psi}D^b\Psi-\Psi\overline{D^b\Psi}).
\end{eqnarray}
By the holographic dictionary, the dual boundary system is placed at a finite temperature given by 
\begin{equation}
T=\frac{3}{4\pi z_h},
\end{equation}
where a conserved current operator $J$ is sourced by the boundary value of the bulk gauge field $A$ and a scalar operator $O$ of conformal dimension $\Delta=\frac{3}{2}\pm\sqrt{\frac{9}{4}+m^2L^2}$ is sourced by the near boundary data of scalar field $\Psi$.  For simplicity but without loss of generality, we shall focus only on the case of $m^2L^2=-2$ in the axial gauge $A_z=0$, in which the asymptotic solution of $A$ and $\Psi$ can be expanded near the AdS boundary as
\begin{equation}\label{near}
A_\nu=a_\nu+b_\nu z+o(z),\quad\Psi=\frac{z}{L}[\phi+\psi z+o(z)].
\end{equation}
Then the expectation value of $J$ and $O$ can be explicitly obtained by holography as the variation of renormalized bulk on-shell action with respect to the sources, i.e., 
\begin{eqnarray}
\langle J^\nu\rangle&=&\frac{\delta S_{ren}}{\delta a_\nu}=\lim_{z\rightarrow 0}\sqrt{-g}F^{z\nu},\label{current}\\
\langle O\rangle&=&\frac{\delta S_{ren}}{\delta \phi}=\bar{\psi}-\dot{\bar{\phi}}-ia_t\bar{\phi},\label{vev}
\end{eqnarray}
where the renormalized action is obtained by adding a counter term to the original action to make it finite as $S_{ren}=S-\frac{1}{L}\int_\mathcal{B}\sqrt{-\gamma}|\Psi|^2$, and the dot denotes the time derivative\cite{LTZ}. When this scalar operator $O$ develops a nonzero expectation value spontaneously in the situation where the source is switched off, the system is driven into a superfluid phase with $\langle O\rangle=\bar{\psi}$ characterizing the superfluid condensate\cite{units}. Generically such a superfluid phase has gapped vortex excitations with the circulation quantized. With the superfluid velocity defined as\cite{ACL}
\begin{equation}
\mathbf{u}=\frac{\mathbf{j}}{|\psi|^2},\quad \mathbf{j}=\frac{i}{2}(\bar{\psi}\mathbf{\partial}\psi-\psi\mathbf{\partial}\bar\psi),
\end{equation}
the winding number $w$ of a vortex is determined by 
\begin{equation}\label{winding}
w=\frac{1}{2\pi}\oint_\gamma d\mathbf{x}\cdot\mathbf{u},
\end{equation}
where $\gamma$ denotes a counterclockwise oriented path surrounding a single vortex. In particular, close to the core of a single vortex with winding number $w$, the condensate $\bar{\psi}\propto(\mathbf{z}-\mathbf{z_0})^w$ for $w>0$ and $\psi\propto(\mathbf{z}-\mathbf{z_0})^{-w}$ for $w<0$ with $\mathbf{z}$ the complex coordinate and $\mathbf{z_0}$ the location of the core. Thus not only does the magnitude of condensate apparently vanish at the core of a vortex but also the corresponding phase shift around the vortex is given precisely by $2\pi w$. This is the characteristic property of a vortex and will be used as an efficient way to identify vortices in our later vortex counting.

To address the vortex pair annihilation in a turbulent superfluid by holography, we would first like to impose the following boundary conditions onto the bulk fields, i.e.,
\begin{equation}\label{sourceless}
\phi=0,\quad a_t=\mu,\quad\mathbf{a}=0,
\end{equation}
where we set the chemical potential $\mu>\mu_c$ with $\mu_c$ the critical chemical potential for the onset of a homogeneous superfluid phase, given by $\mu_c\approx4.07$. Explicit gravity solutions dual to a static vortex of arbitrary winding number have been numerically constructed in \cite{AJ,MPS,KKNY}. But we are required to prepare an initial bulk configuration for $\Psi$ at the Eddington time $t=0$ such that the dual initial boundary state includes 300 vortex-antivortex pairs in a $100\times 100$ square box with periodic boundary conditions, where the vortices (each with winding number $w=1$) and antivortices (each with winding number $w=-1$) are randomly placed, mimicking the experimental setup\cite{KMCSS}.

{The detailed construction of the initial bulk configuration for $\Psi$ is basically similar to that in the Appendix of Ref.\cite{ACL}, with the main difference that our vortices are placed randomly instead of on a lattice. Due to the asymptotic behavior (\ref{near}), we define $\Phi:=\frac{\Psi}{z}$ as usual for our numerical convenience, and the homogeneous equilibrium  configuration $\Phi_\mathrm{eq}(z)$ can be easily obtained as in the ordinary holographic superfluid. For the single vortex configuration $\Phi_w=g(r)\Phi_\mathrm{eq}(z)e^{i w\theta}$ ($w=\pm1$) in polar coordinates $(r,\theta)$, which is used to compose the initial state here, the concrete form of the profile function $g(r)$ is unimportant. Actually, besides $g(r)\to 1$ for $r\to\infty$, the only requirement is $g(r)\propto r$ for $r\to 0$, which guarantees the smoothness of $\Phi_w(r\to 0)$. In practice, $g(r)=\tanh(r/c)$ with the constant $c\sim\mathcal{O}(1)$ just works, and we have checked that there is no observable effect on our results by different choices of $c$ or different forms of $g(r)$.}

Then the initial data of $A_t$ can be determined by the constraint equation
\begin{equation}\label{constraint}
\partial_z(\partial_zA_t-\mathbf{\partial}\cdot\mathbf{A})=i(\bar{\Phi}\partial_z\Phi-\Phi\partial_z\bar{\Phi})
\end{equation}
once $\mathbf{A}$ is given at $t=0$ supplemented by the second boundary condition $\partial_z A_t|_{z=0}=-\rho$ with $\rho$ the boundary charge density, in addition to $a_t=A_t|_{z=0}=\mu$
%, instead of setting $A_t|_{z=z_h}=0$ as in \cite{ACL}
%\footnote{As explained below and in the Appendix, it is crucial to choose $\rho$ as the controlling parameter of the evolution in our study, which will investigate the physics at various given (equilibrium) chemical potential $\mu$. There is no real problem at this point in \cite{ACL}, since it is the scaling behavior and energy cascade that is considered there and the final equilibrium chemical potential is not important.}
. For convenience but without loss of generality, we shall let the initial value $\mathbf{A}=0$. Importantly, the initial charge density $\rho$ is taken to be that for the homogeneous and isotropic superfluid phase corresponding to the chemical potential $\mu$, the reason of which is explained in detail in the Appendix. With the above initial data and boundary conditions, the later time behavior of bulk fields can be obtained by the following evolution equations
\begin{eqnarray}\label{evolution1}
\partial_t\partial_z\Phi &=& iA_t\partial_z\Phi+\frac{1}{2}[i\partial_zA_t\Phi+f\partial_z^2\Phi+f^\prime\partial_z\Phi\nonumber\\
&&+(\mathbf{\partial}-i\mathbf{A})^{2}\Phi-z\Phi],\\
\label{evolution2} \partial_t\partial_z\mathbf{A} &=& \frac{1}{2}[\partial_z(\mathbf{\partial}A_t+f\partial_z\mathbf{A})+(\mathbf{\partial}^2\mathbf{A}-\mathbf{\partial}\mathbf{\partial}\cdot\mathbf{A})\nonumber\\
&&-i(\bar{\Phi}\mathbf{\partial}\Phi-\Phi\mathbf{\partial}\bar{\Phi})]-\mathbf{A}\bar{\Phi}\Phi, \\
\label{evolution3}    \partial_t\partial_zA_t &=& \mathbf{\partial}^2A_t+f\partial_{z}\mathbf{\partial}\cdot\mathbf{A}-\partial_t\mathbf{\partial}\cdot\mathbf{A}-2A_t\bar{\Phi}\Phi \nonumber\\
&& +i f(\bar{\Phi}\partial_{z}\Phi-\Phi\partial_{z}\bar{\Phi})-i(\bar{\Phi}\partial_t\Phi-\Phi\partial_t\bar{\Phi}). 
\end{eqnarray}
We numerically solve these non-trivial evolution equations by employing pseudo-spectral methods plus Runge-Kutta method. Namely, we expand all the involved bulk fields in a basis of Chebyshev polynomials in the $z$ direction as well as Fourier series in the $\mathbf{x}$ direction, and plug such expansions into the above (3+1)D partial differential equations to make them boil down into a set of 1D ordinary differential equations, which is well amenable to the time evolution with the fourth order Runge-Kutta scheme. We also use the constraint equation (\ref{constraint}) to check the validity of our resultant numerical solution.

The above numerical evolution scheme has the drawback that the numerical violation of the constraint equation (\ref{constraint}) is accumulative in time, so in principle it is not suitable for long time evolution, though we have carefully made the violation under control by appropriately choosing the numbers of the Chebyshev and Fourier expansion modes as well as the time step of the Runge-Kutta method. Alternatively, we can use another numerical evolution scheme as described in the Appendix, which has a constraint violation accumulated in the spatial direction $z$ instead of time and is believed to be more suitable for long time evolution than the scheme described above. In fact, both these schemes have been used in our study to achieve double confidence.

Finally, the vortex dynamics can be decoded by extracting the near boundary behavior of $\Psi$ according to (\ref{near}), (\ref{vev}) and (\ref{sourceless}), i.e. the condensate $\langle O\rangle=\bar{\psi}$ is just conjugate of the boundary ($z=0$) value of $\Phi$. For identifying vortices, we calculate winding numbers (\ref{winding}) by computing the total phase difference of $\psi$ around each plaquette formed by the nearest neighboring collocation points in the pseudo-spectral Fourier collocation on the $z=0$ surface.
\begin{figure}
\begin{center}
\includegraphics[width=3.5cm,bb=0 0 800 800]{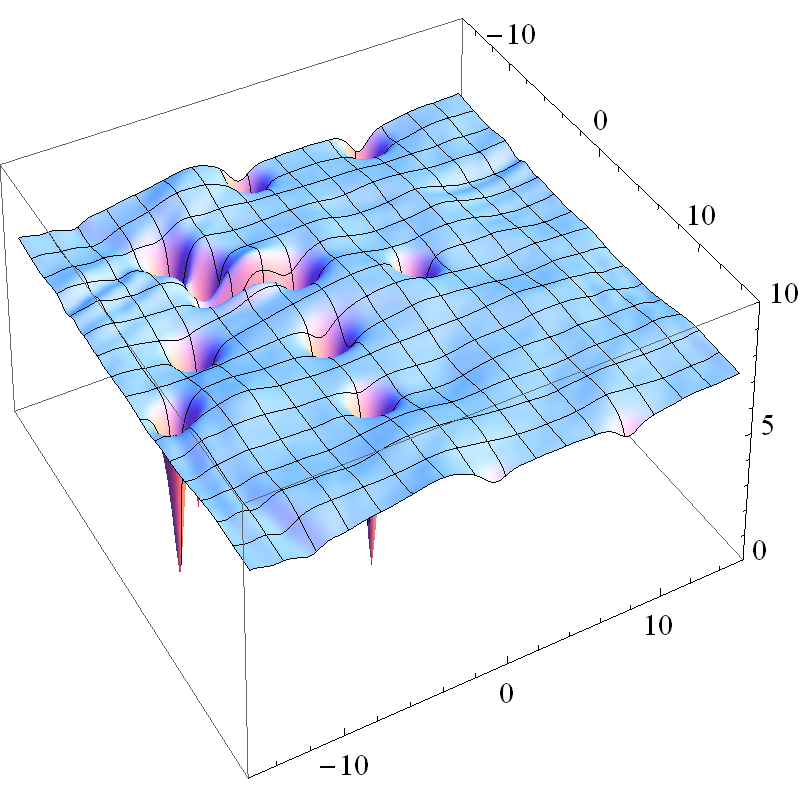}
\includegraphics[width=3.5cm,bb=0 0 800 800]{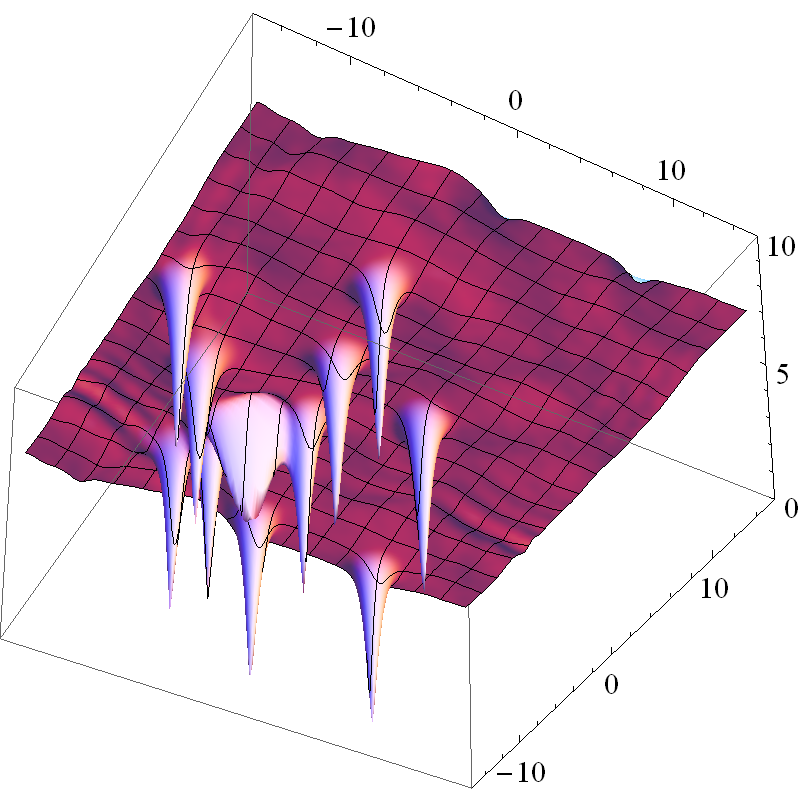}
\includegraphics[width=3.5cm,bb=0 0 801 802]{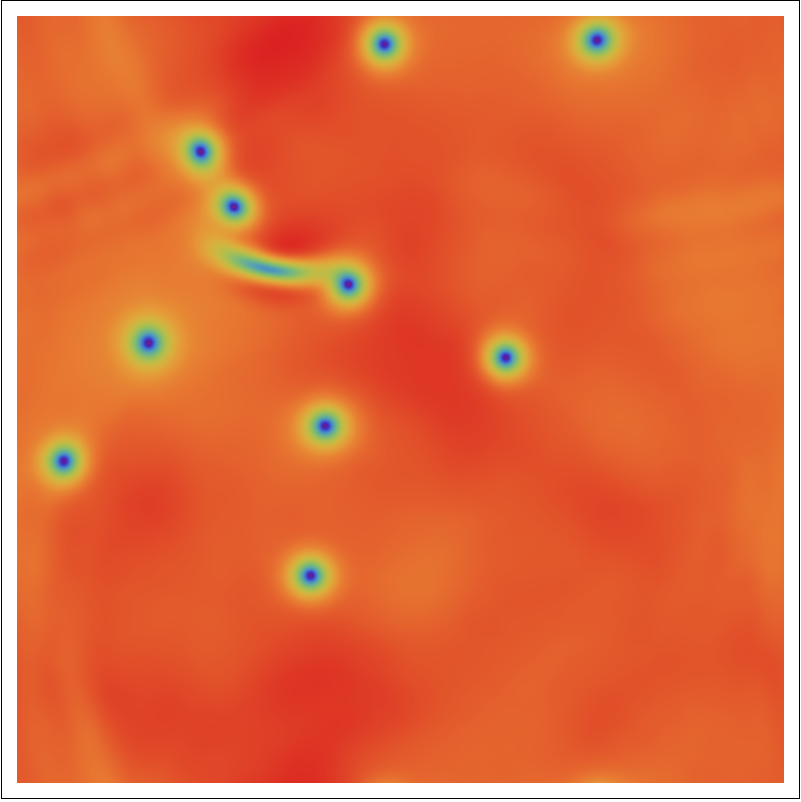}
\includegraphics[width=3.5cm,bb=0 0 880 881]{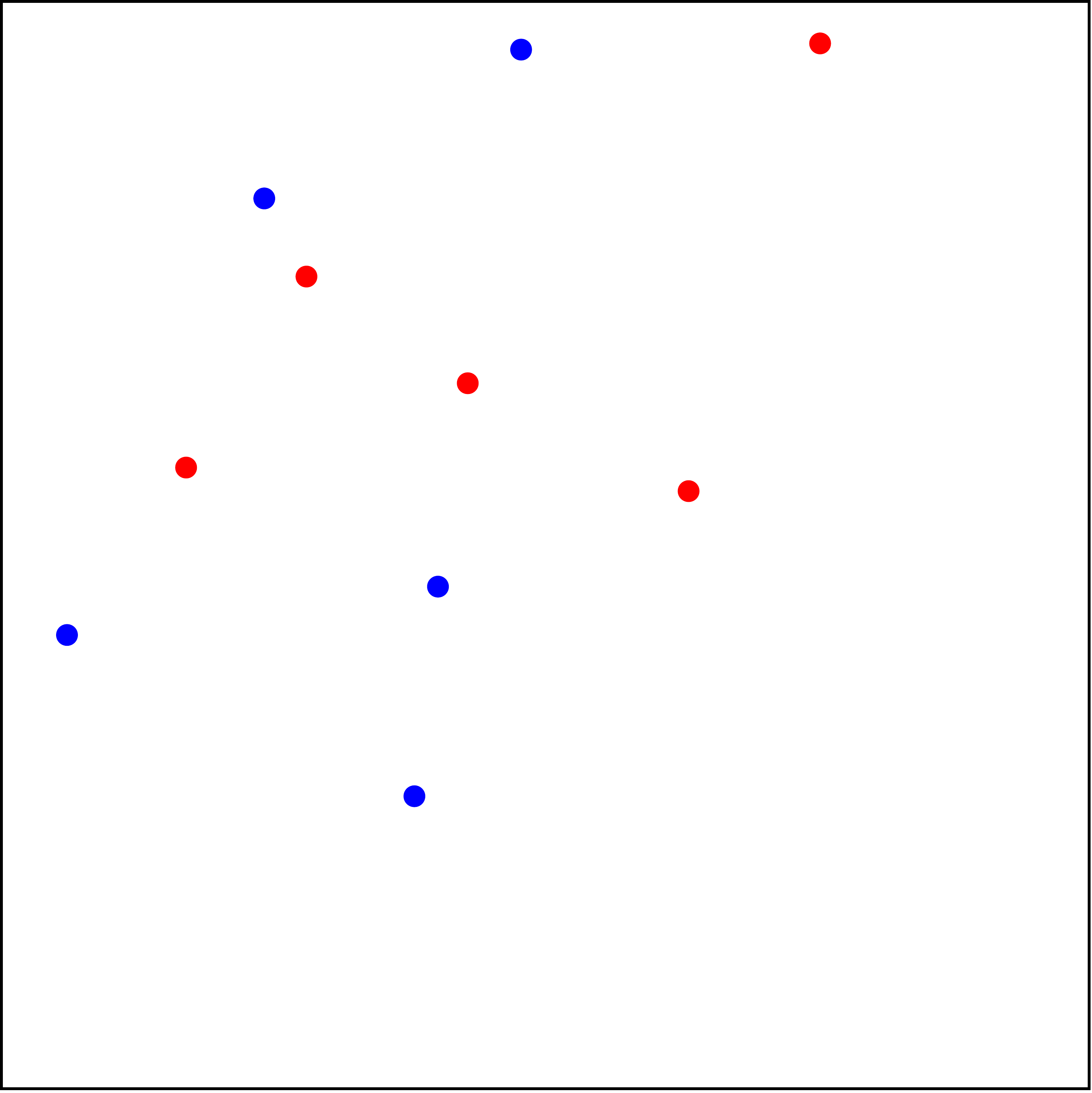}
\includegraphics[width=3.5cm,bb=0 0 800 800]{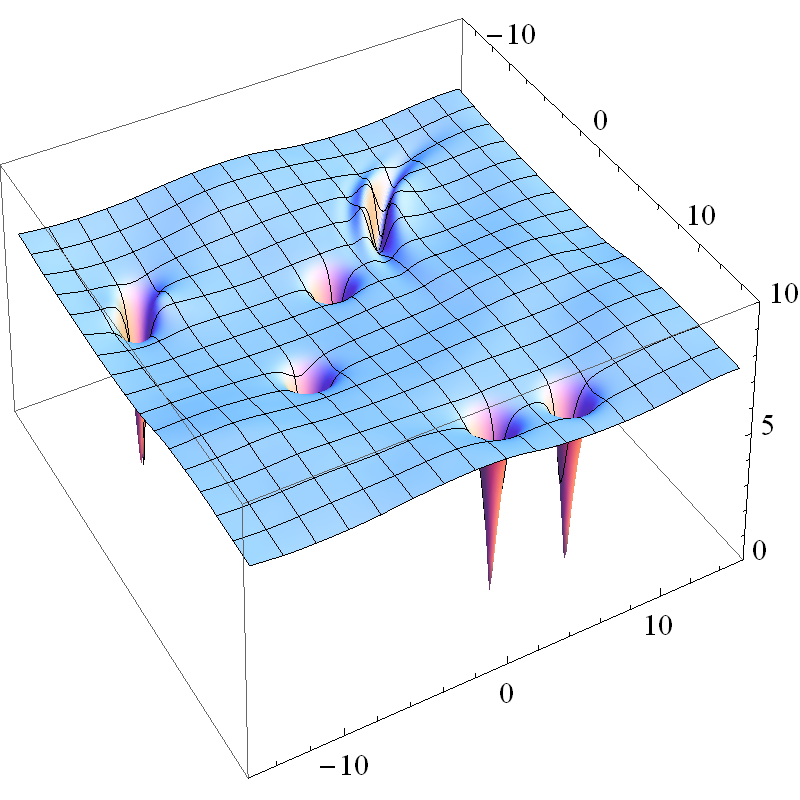}
\includegraphics[width=3.5cm,bb=0 0 800 800]{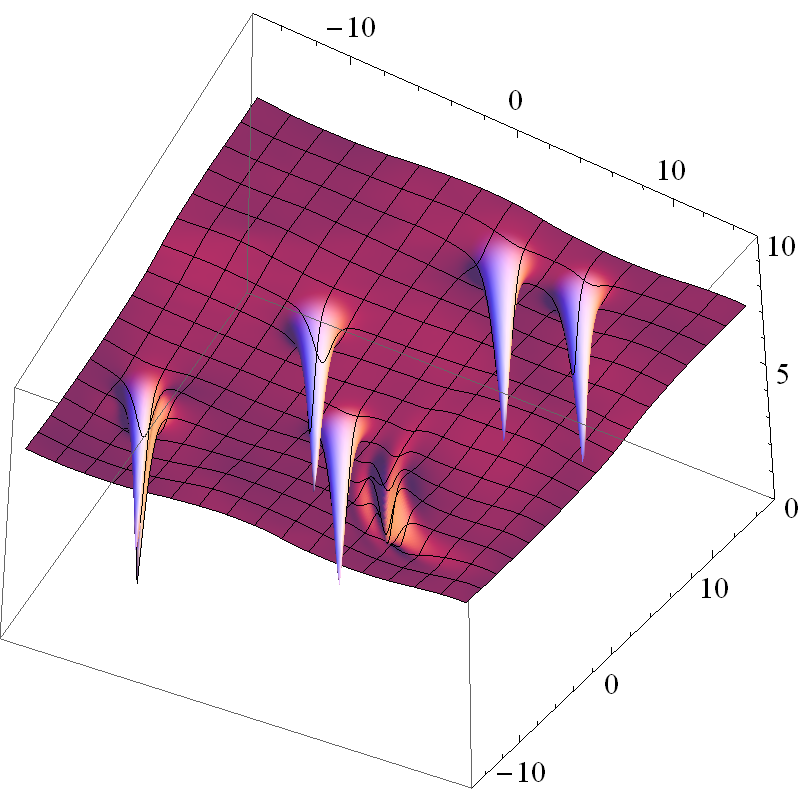}
\includegraphics[width=3.5cm,bb=0 0 801 802]{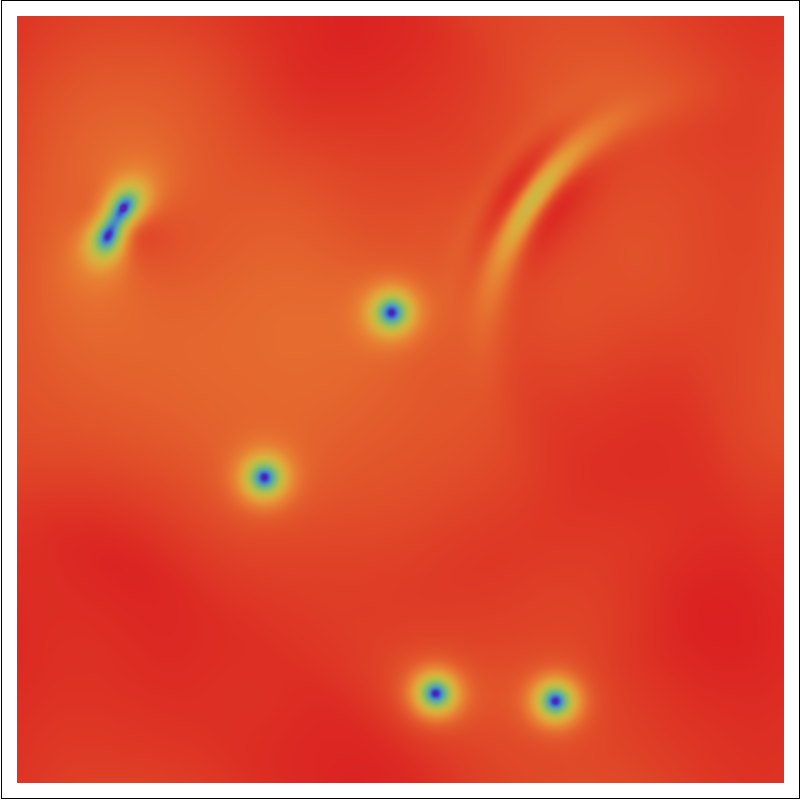}
\includegraphics[width=3.5cm,bb=0 0 880 881]{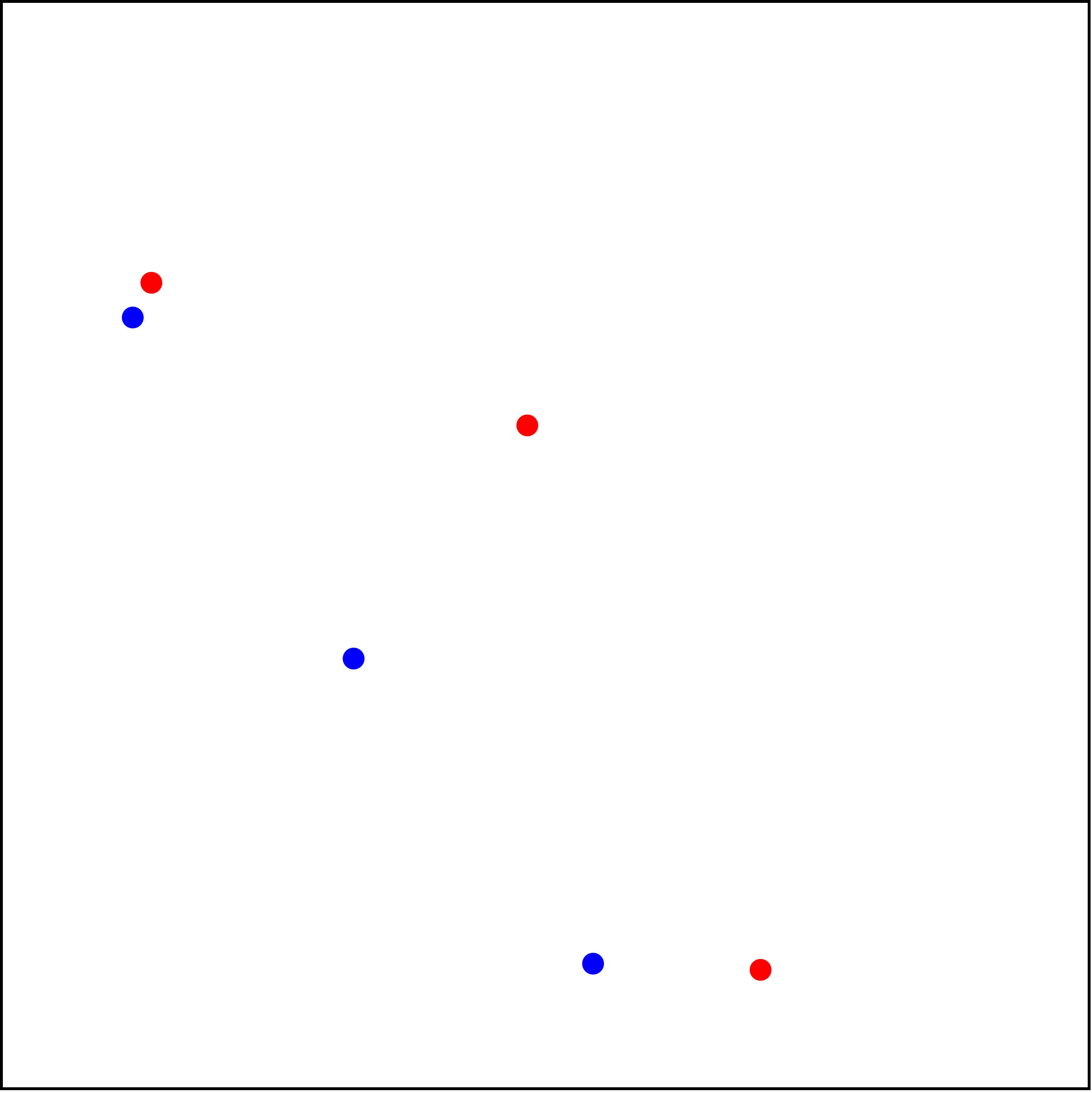}
\end{center}
\caption{The snapshots of the turbulent superfluid at the chemical potential $\mu=6.25$, where the top is for  $t=100$ and the bottom is for $t=200$.  The first two columns are 3D plots of the absolute value of condensate, and the third column is the density plot of the absolute value of condensate. In addition,  in the last column the blue and red points denote the core positions of vortices and antivortices respectively. The shock waves are seen as the ripples, the gray solitons are identified in the form of the bending structure with the condensate depleted, and the vortex cores are located at the position where the condensate vanishes.}
\label{updown}
\end{figure}

\section{Numerical results}
We now describe the typical behaviors in the holographic turbulent superfluid constructed above by numerically solving the bulk equations of motion for a variety of random initial conditions at each chemical potential we choose. As time passes, we never see the merging of vortices with the same circulation. Instead, we observe that the coalescence of vortex and antivortex cores is followed by formation of a crescent-shaped gray soliton when the size of the vortex dipole becomes smaller than a certain threshold value $d$. Such a crescent-shaped gray solitons originates in the fact that the coalesced vortex and antivortex cores generically march forward leading to a perpendicular linear momentum of the vortex dipole to the vortex dipole direction, and converts eventually into a shock wave, dissipatively propagating in the superfluid. With such vortex pair annihilation featured process, the vortex number decreases and eventually the turbulent condensate relaxes into a homogeneous and isotropic equilibrium state. For the purpose of demonstration, we plot one early time and one late time configurations of turbulent superfluid at the chemical potential $\mu=6.25$ respectively in FIG.\ref{updown}, where 30 vortex-antivortex pairs are prepared in a $30\times 30$ periodic square box as the initial state\cite{video}. It is noted that the crescent shape of the gray solitons in our simulation is remarkably similar to that of the density-depleted regions observed in the experiment (see Fig.3 in \cite{KMCSS}).
\begin{figure}
\begin{center}
\includegraphics[width=7.0cm,bb=0 0 432 494]{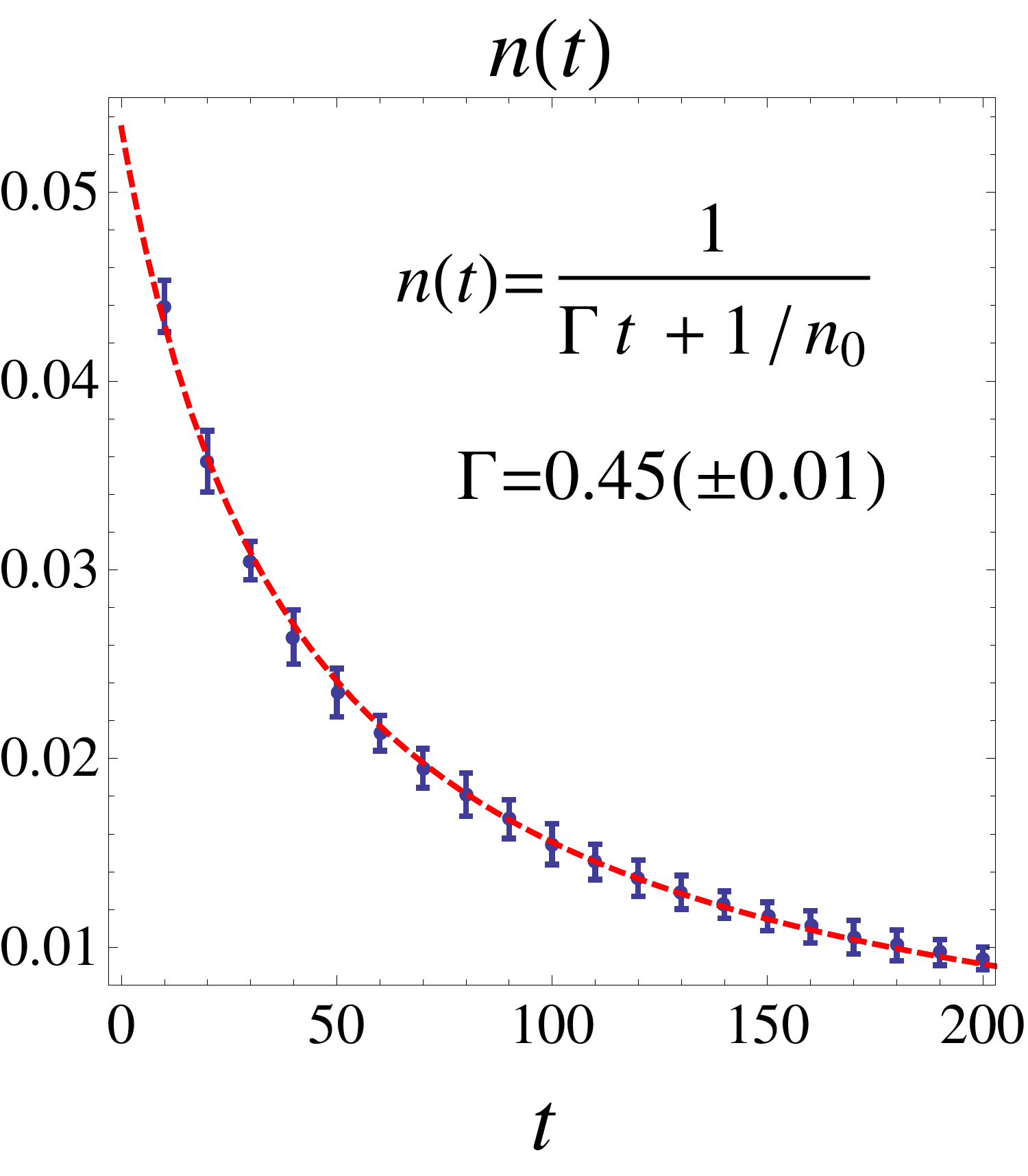}
\includegraphics[width=7.0cm,bb=0 0 432 502]{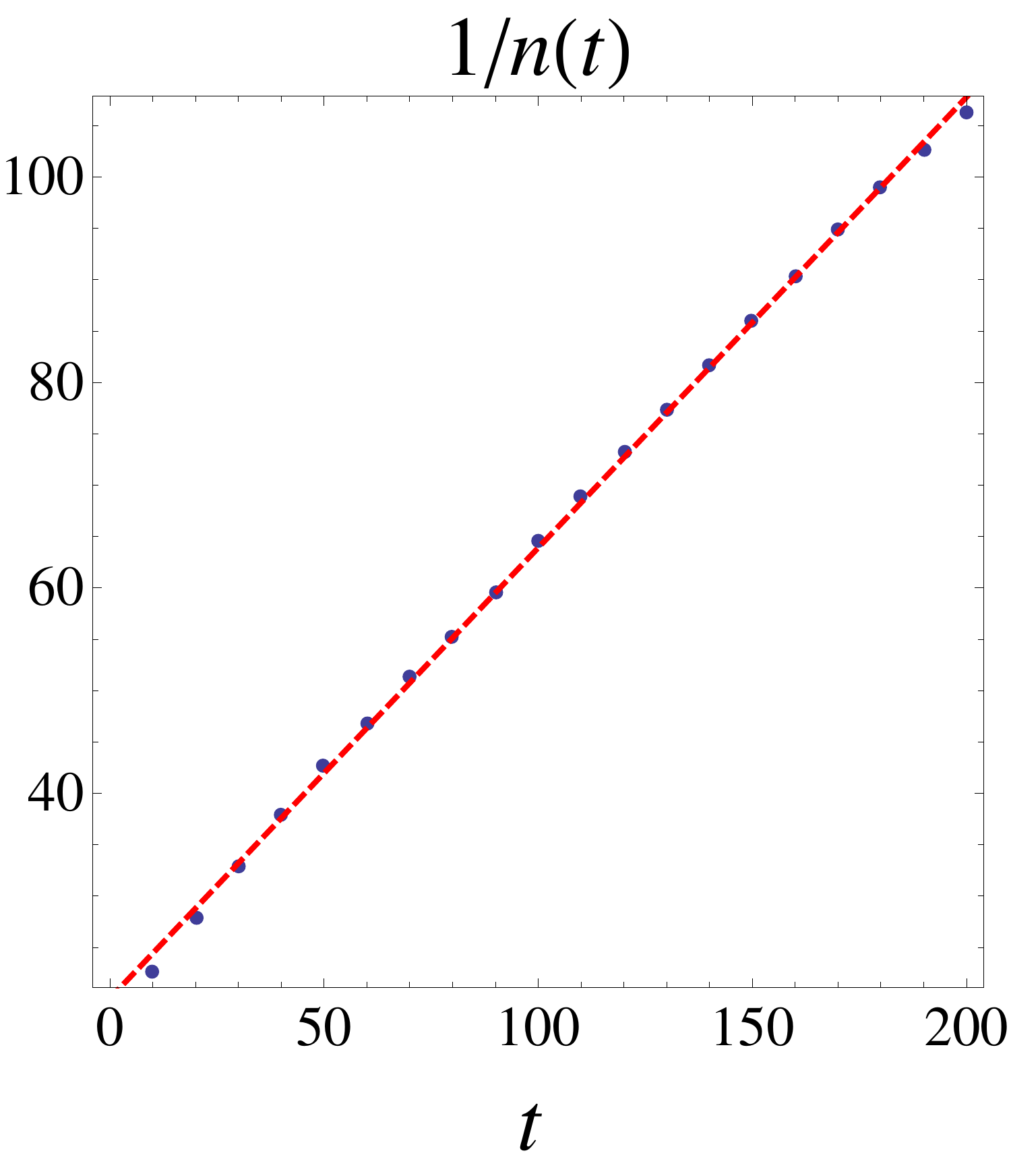}
\end{center}
\caption{The temporal evolution of averaged vortex number density in the turbulent superfluid over 12 groups of data with randomly prepared initial conditions at the chemical potential $\mu=6.25$, which is well described by the formula (\ref{decayrate}), as implied by vortex pair annihilation mechanism.}
\label{datafit}
\end{figure}

Instead of the energy spectrum investigated in \cite{ACL}, we shall focus on the quantitative behavior for the temporal evolution of vortex number in the above relaxation process because this behavior has already been accessible experimentally\cite{KMCSS}.  Note that it takes $N$ vortices to find $N$ antivortices, thus it is reasonable for one to expect  that the annihilation rate should be proportional to $N\times N = N^2$. In terms of the number density, the vortex decay takes the following form
\begin{equation}
\frac{dn(t)}{dt}=-\Gamma n(t)^2,
\end{equation}
whereby one can obtain
\begin{equation}\label{decayrate}
n(t)=\frac{1}{\Gamma t+\frac{1}{n_0}},
\end{equation}  
where the decay rate is suggested by the kinetic theory to be proportional to the product of the velocity and cross section of vortices, namely $\Gamma=\frac{vd}{2}$ with $v$ the velocity of vortices if the vortices can be regarded as a gas of particles. Thus the decay rate is supposed to be uniquely determined by the chemical potential through $v$ and $d$. On the other hand, it is important to focus on the statistical laws because the driven turbulent flow is chaotically sensitive to our randomly prepared initial conditions. Therefore we run $12$ groups of data for each chemical potential and extract the corresponding decay rate by fitting the temporal evolution of the averaged number density with the statistical error by the formula (\ref{decayrate}). As a demonstration, we only plot the relevant result for the case of chemical potential $\mu=6.25$ in FIG.\ref{datafit}. Obviously, the decrease of vortex number density is well captured by vortex pair annihilation induced two-body decay as (\ref{decayrate}) from a very early time on. The similar results are also obtained for other chemical potentials. It is noteworthy that although we here focus on the early time evolution such a decay pattern is believed to persist towards a very late time\cite{EGKS,Samberg}.
The upshot is that because the drift-out effect is favorably absent from our holographic superfluid due to the periodic boundary conditions imposed on the square box, the above results inarguably confirms the suspected two-body decay mechanism by vortex pair annihilation in \cite{KMCSS}.

\begin{figure}
\begin{center}
\includegraphics[width=8.0cm,bb=0 0 432 312]{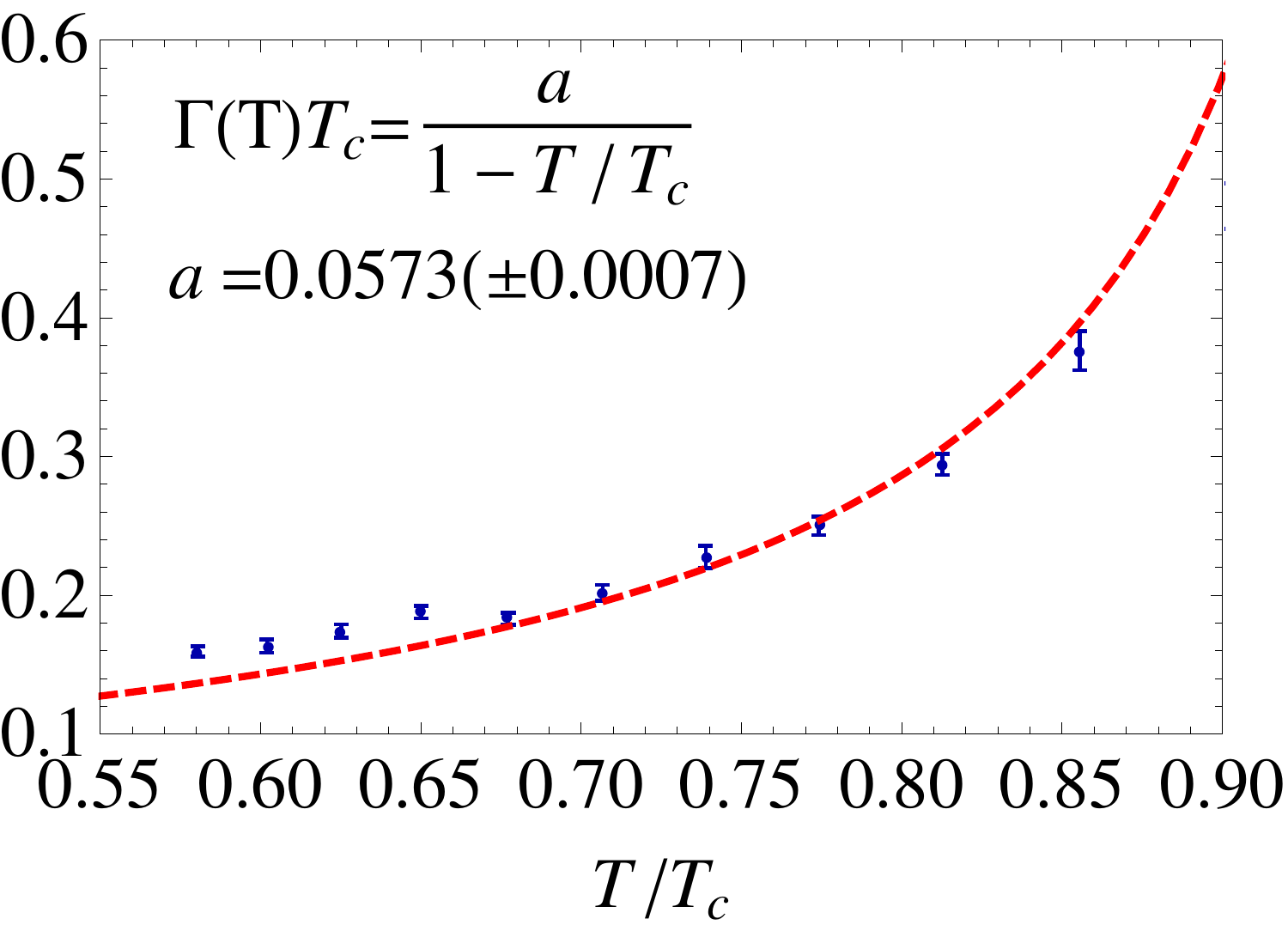}
\end{center}
\caption{The variation of decay rate with respect to the temperature, which is obtained by the scaling symmetry of our theory from the data for those equally spaced chemical potentials from $\mu=4.75$ to $\mu=7.0$. The decay rate is further fit by the formula (\ref{mean}) for the six data points nearest to the critical temperature (the rightmost points).}
\label{statlaw}
\end{figure}

We further plot the variation of decay rate with respect to the temperature in FIG.\ref{statlaw} by the scaling symmetry of our theory. As illustrated, the decay rate is increased with the temperature within the error bars. In particular, the decay rate is expected to be divergent when one approaches the critical point. Inspired by the effective description of 2D superfluid turbulence in which the decay rate is expected to be proportional to the inverse of superfluid density $n_s$\cite{CL,Chesler}, we further fit the near critical point data by the following formula
\begin{equation}\label{mean}
\Gamma(T)T_c=\frac{a}{1-\frac{T}{T_c}}
\end{equation}  
with ${T_c}$ the critical temperature because $n_s=|\langle O\rangle|^2\propto 1-\frac{T}{T_c}$ near the critical point, which is typical of second order phase transitions\cite{HHH1}. As one can see, the data roughly confirm the above ansatz for the temperature dependence of decay rate in (\ref{mean}). 

We would like to end this section by mentioning the near zero temperature behavior of decay rate. Actually the above effective description of 2D superfluid turbulence is further indicative of the low temperature decay rate $\Gamma\propto T^2$ by using the very fact that the force on a moving vortex at low temperature can be expressed in terms of Kubo formulas of defect CFT operators because at low energies the vortex in a holographic superfluid can be viewed as a conformal defect, with a CFT$_1$ living on it\cite{DHIS}.  Although as shown in \cite{HHH2}, the probe limit we are working with can capture accurately the essential physics all the way down to zero temperature, the numerical simulation turns out to be too time consuming to be worked out by our limited computational resources for the vortex dynamics at low temperature. But we hope to report it elsewhere in the future. 

\section{Conclusion and discussion}
The superfluid dynamics at zero temperature is generally described with the Gross-Pitaevskii equation. But in order to address the finite temperature superfluid dynamics, the dissipative terms are usually introduced in a purely phenomenological way. On the contrary, holographic duality, as a new laboratory and powerful tool, offers a first-principles method to study vortex dynamics in the turbulent superfluid, where the superfluid at finite temperature is dual to a hairy black hole in the bulk and the dissipation mechanism is universally captured by excitations absorbed by the hairy black hole.

We use this gravitational description to numerically construct turbulent non-counterflows with the initial vortices and antivortices placed randomly in the 2D finite temperature holographic superfluid. We find that in the thermal relaxation process the decrease of the vortex number obeys the intrinsic two-body decay due to the vortex pair annihilation, thus confirm the recent experiment data based suspicion. Furthermore, near the critical temperature such a decay pattern is in good agreement with the recently developed effective theory of 2D superfluid turbulence.

It is worthwhile mentioning the relevant result obtained by the effective field theory of superfluid dynamics in \cite{CL}, where the $N^2$ decay behavior is found for the laminar flow while $N^\frac{5}{3}$ decay behavior for the turbulent flow. This result appears to be at odds with the $N^2$ decay behavior for our turbulent flow. But the point is that we are dealing with the two different regimes of superfluid dynamics. To be more precise, Ref. \cite{CL} is dealing with the superfluid at a very low temperature with a very tiny dissipative term  and dilute vortices while we are dealing with the superfluid  at the temperature order of critical temperature where the dissipation is supposed to be very large and the vortices are not necessarily dilute. So there is no reason to claim that the superfluid dynamics should exhibit the similar behavior at these two regimes. Actually the inverse energy cascade is obeyed by the turbulent flow in \cite{CL} while as shown in \cite{ACL} our superfluid turbulence exhibits a direct energy cascade. Therefore it is highly interesting to see whether there is a critical temperature for our superfluid to transition from the direct energy cascade $N^2$ behavior to the inverse energy cascade $N^\frac{5}{3}$ behavior, albeit beyond the scope of our paper.

In addition, in order for more detailed comparison between the holographic simulation and the experiments, more works should be done on both sides. As mentioned previously, at very low temperature the current numerical approach tends to break down even in the probe limit, since both the equilibrium configuration in the $z$ direction and the vortex configuration in the radial direction (with respect to the vortex center) tend to be non-analytic, so a very large number of expansion modes should be taken in order to maintain precision, which challenges any computational resources. Possibly novel approach should be used to overcome this difficulty. On the experimental side, after all, the highly oblate Bose-Einstein condensate is not a genuine 2D system\cite{KMCSS}. It is also difficult (if ever possible) to properly subtract the drift-out effect, which seems inevitable in experimental setups. This fact makes the existing experimental data insufficient to determine the intrinsic physics of vortex pair annihilation. In addition, the experimental data at very low temperature or near criticality is also lacking. We hope that significant progresses will be achieved on both the theoretical (numerical) and experimental sides in the near future.

\section*{Acknowledgements}
We thank Yong-il Shin for his stimulating talk and later helpful discussion on vortex pair annihilation on the focus conference ``Precision Tests of Many-Body Physics with Ultracold Quantum Gases" during the long program ``Precision Many-Body Physics of Strong Correlated Quantum Matter" at KITPC. 
We would like to express a special thanks to Hong Liu for his insightful comments and valuable suggestions throughout this whole project.   
We acknowledge the organizers of another long term program ``Quantum Gravity, Black Holes and Strings" at KITPC for the fantastic infrastructure they provide and the generous financial support they offer.
H.Z. is grateful to the Mainz Institute for Theoretical Physics for its hospitality and its partial support during his attending the program ``String Theory and its Applications", and CERN for its financial support during his attending ``CERN-CKC TH Institute on Numerical Holography",  where he benefits much from the discussions with Andreas Samberg. He also acknowledges the Erwin Schr\"{o}dinger Institute for Mathematical Physics for the financial support during his participation in the program ``Topological Phases of Quantum Matter", where the relevant conversation with Michael Stone is much appreciated. Finally, H.Z. thanks the Galileo Galilei Institute for Theoretical Physics for the hospitality and the INFN for partial support during his attending the workshop ``Holographic Methods for Strongly Coupled Systems", where the revision of this work is being conducted.
Y.D. and Y.T. are partially supported by NSFC with Grant No.11475179. C.N. is partially supported by NSFC with Grant No.11275208. H.Z. is supported in part by the Belgian Federal Science Policy Office through the Interuniversity Attraction Pole P7/37, by FWO-Vlaanderen through the project 
G020714N, and by the Vrije Universiteit Brussel through the
Strategic Research Program ``High-Energy Physics". He is also an individual FWO fellow supported by 12G3515N. 

\section*{Appendix: More on our numerical scheme}
Besides the numerical scheme described in the main body of this paper, where the constraint equation (\ref{constraint}) is imposed at $t=0$, and used as a later monitor of the reliability of our numerical evolution by the equations (\ref{evolution1}), (\ref{evolution2}), and (\ref{evolution3}), we have an alternative scheme described as follows.

In this scheme, we still evolve the scalar field $\Phi$ and the gauge field $\mathbf{A}$
in time by (\ref{evolution1}) and (\ref{evolution2}), respectively, with the fourth order Runge-Kutta method. However, we no longer use
the equation (\ref{evolution3}) to achieve the time evolution of $A_{t}$ except at $z=0$, where (\ref{evolution3}) boils down into
\[
\partial_{t}\partial_{z}A_{t}|_{z=0}=\partial_{z}\partial\cdot\mathbf{A}|_{z=0},
\]
which is nothing but the conservation equation for the boundary global current (\ref{current}) in the boundary system ($z=0$). Upon integration on the constant time surface (within the boundary $z=0$), this current gives the conserved charge during the dynamical evolution process, which guarantees that the boundary system will eventually settle down to the desired chemical potential $\mu$ if we use the corresponding (equilibrium) charge density $\rho$ as the initial charge density.
Then with the solution for $\partial_{z}A_{t}(t)|_{z=0}$ as well as $A_{t}(t)|_{z=0}=\mu$,
we shall instead use the constraint equation (11) to solve $A_{t}$ from $\Phi$ and $\mathbf{A}$ not only at the initial time $t=0$ but also at every
step of the later time evolution. It is not hard to show that 
the equation (\ref{evolution3}) will automatically hold elsewhere by virtue of the equations
(\ref{constraint}), (\ref{evolution1}), and (\ref{evolution2}), if it does hold at a given constant $z$ hypersurface. 

In both schemes, we take
28 Chebyshev modes in the $z$ direction and 361 Fourier modes in
both the $x$ and $y$ directions. The advantage of this alternative scheme is that it can undergo a very long time evolution because the main numerical error comes from the
violation of the equation (\ref{evolution3}) deep into the bulk horizon, which is nevertheless not accumulated fast with the time evolution. For
the chemical potential we have chosen, the violation of the equation (\ref{evolution3}) at
the horizon $z=1$ is of the order of $10^{-4}$, which we deem to be acceptable. Regarding the main results presented in this paper, both schemes are in good agreement with each other, which can be regarded as a double check of the reliability of our results.

{For the case of the smallest size of a vortex core ($\mu=7.0$), there are roughly 20 grids within each vortex core, which is expected to be fine enough. For the time step $\Delta t$ in the Runge-Kutta evolution, we have taken both $\Delta t=0.05$ and $\Delta t=0.025$, and find no observable difference for a given initial state on either the vortex number counting (to $t=200$) or the constraint violation as described above. For $\Delta t=0.025$, it takes about two weeks on a mainstream desktop computer for one run (to $t=200$). Actually, we have 10 different chemical potentials and for each chemical potential we need 12 groups of data, which are accomplished on a parallel work station.}


\begin{thebibliography}{99}
\bibitem{large} For notational convenience we shall set the overall factor $16\pi G q^2=1$ later on.
\bibitem{units}In what follows, we shall work in the units where $L=1$ and $z_h=1$.
\end{thebibliography}
\end{document}